\documentclass[letterpaper,compsoc,twoside]{IEEEtran}
\usepackage{fixltx2e} 
\usepackage{cmap} 
\usepackage{ifthen}
\usepackage[T1]{fontenc}
\usepackage[utf8]{inputenc}

\usepackage[font={small,it},labelfont=bf]{caption}
\usepackage{float}

\pdfoutput=1
\usepackage{scipy}
\makeatletter
\def\PY@reset{\let\PY@it=\relax \let\PY@bf=\relax%
    \let\PY@ul=\relax \let\PY@tc=\relax%
    \let\PY@bc=\relax \let\PY@ff=\relax}
\def\PY@tok#1{\csname PY@tok@#1\endcsname}
\def\PY@toks#1+{\ifx\relax#1\empty\else%
    \PY@tok{#1}\expandafter\PY@toks\fi}
\def\PY@do#1{\PY@bc{\PY@tc{\PY@ul{%
    \PY@it{\PY@bf{\PY@ff{#1}}}}}}}
\def\PY#1#2{\PY@reset\PY@toks#1+\relax+\PY@do{#2}}

\expandafter\def\csname PY@tok@gd\endcsname{\def\PY@tc##1{\textcolor[rgb]{0.63,0.00,0.00}{##1}}}
\expandafter\def\csname PY@tok@gu\endcsname{\let\PY@bf=\textbf\def\PY@tc##1{\textcolor[rgb]{0.50,0.00,0.50}{##1}}}
\expandafter\def\csname PY@tok@gt\endcsname{\def\PY@tc##1{\textcolor[rgb]{0.00,0.27,0.87}{##1}}}
\expandafter\def\csname PY@tok@gs\endcsname{\let\PY@bf=\textbf}
\expandafter\def\csname PY@tok@gr\endcsname{\def\PY@tc##1{\textcolor[rgb]{1.00,0.00,0.00}{##1}}}
\expandafter\def\csname PY@tok@cm\endcsname{\let\PY@it=\textit\def\PY@tc##1{\textcolor[rgb]{0.25,0.50,0.56}{##1}}}
\expandafter\def\csname PY@tok@vg\endcsname{\def\PY@tc##1{\textcolor[rgb]{0.73,0.38,0.84}{##1}}}
\expandafter\def\csname PY@tok@m\endcsname{\def\PY@tc##1{\textcolor[rgb]{0.13,0.50,0.31}{##1}}}
\expandafter\def\csname PY@tok@mh\endcsname{\def\PY@tc##1{\textcolor[rgb]{0.13,0.50,0.31}{##1}}}
\expandafter\def\csname PY@tok@cs\endcsname{\def\PY@tc##1{\textcolor[rgb]{0.25,0.50,0.56}{##1}}\def\PY@bc##1{\setlength{\fboxsep}{0pt}\colorbox[rgb]{1.00,0.94,0.94}{\strut ##1}}}
\expandafter\def\csname PY@tok@ge\endcsname{\let\PY@it=\textit}
\expandafter\def\csname PY@tok@vc\endcsname{\def\PY@tc##1{\textcolor[rgb]{0.73,0.38,0.84}{##1}}}
\expandafter\def\csname PY@tok@il\endcsname{\def\PY@tc##1{\textcolor[rgb]{0.13,0.50,0.31}{##1}}}
\expandafter\def\csname PY@tok@go\endcsname{\def\PY@tc##1{\textcolor[rgb]{0.20,0.20,0.20}{##1}}}
\expandafter\def\csname PY@tok@cp\endcsname{\def\PY@tc##1{\textcolor[rgb]{0.00,0.44,0.13}{##1}}}
\expandafter\def\csname PY@tok@gi\endcsname{\def\PY@tc##1{\textcolor[rgb]{0.00,0.63,0.00}{##1}}}
\expandafter\def\csname PY@tok@gh\endcsname{\let\PY@bf=\textbf\def\PY@tc##1{\textcolor[rgb]{0.00,0.00,0.50}{##1}}}
\expandafter\def\csname PY@tok@ni\endcsname{\let\PY@bf=\textbf\def\PY@tc##1{\textcolor[rgb]{0.84,0.33,0.22}{##1}}}
\expandafter\def\csname PY@tok@nl\endcsname{\let\PY@bf=\textbf\def\PY@tc##1{\textcolor[rgb]{0.00,0.13,0.44}{##1}}}
\expandafter\def\csname PY@tok@nn\endcsname{\let\PY@bf=\textbf\def\PY@tc##1{\textcolor[rgb]{0.05,0.52,0.71}{##1}}}
\expandafter\def\csname PY@tok@no\endcsname{\def\PY@tc##1{\textcolor[rgb]{0.38,0.68,0.84}{##1}}}
\expandafter\def\csname PY@tok@na\endcsname{\def\PY@tc##1{\textcolor[rgb]{0.25,0.44,0.63}{##1}}}
\expandafter\def\csname PY@tok@nb\endcsname{\def\PY@tc##1{\textcolor[rgb]{0.00,0.44,0.13}{##1}}}
\expandafter\def\csname PY@tok@nc\endcsname{\let\PY@bf=\textbf\def\PY@tc##1{\textcolor[rgb]{0.05,0.52,0.71}{##1}}}
\expandafter\def\csname PY@tok@nd\endcsname{\let\PY@bf=\textbf\def\PY@tc##1{\textcolor[rgb]{0.33,0.33,0.33}{##1}}}
\expandafter\def\csname PY@tok@ne\endcsname{\def\PY@tc##1{\textcolor[rgb]{0.00,0.44,0.13}{##1}}}
\expandafter\def\csname PY@tok@nf\endcsname{\def\PY@tc##1{\textcolor[rgb]{0.02,0.16,0.49}{##1}}}
\expandafter\def\csname PY@tok@si\endcsname{\let\PY@it=\textit\def\PY@tc##1{\textcolor[rgb]{0.44,0.63,0.82}{##1}}}
\expandafter\def\csname PY@tok@s2\endcsname{\def\PY@tc##1{\textcolor[rgb]{0.25,0.44,0.63}{##1}}}
\expandafter\def\csname PY@tok@vi\endcsname{\def\PY@tc##1{\textcolor[rgb]{0.73,0.38,0.84}{##1}}}
\expandafter\def\csname PY@tok@nt\endcsname{\let\PY@bf=\textbf\def\PY@tc##1{\textcolor[rgb]{0.02,0.16,0.45}{##1}}}
\expandafter\def\csname PY@tok@nv\endcsname{\def\PY@tc##1{\textcolor[rgb]{0.73,0.38,0.84}{##1}}}
\expandafter\def\csname PY@tok@s1\endcsname{\def\PY@tc##1{\textcolor[rgb]{0.25,0.44,0.63}{##1}}}
\expandafter\def\csname PY@tok@gp\endcsname{\let\PY@bf=\textbf\def\PY@tc##1{\textcolor[rgb]{0.78,0.36,0.04}{##1}}}
\expandafter\def\csname PY@tok@sh\endcsname{\def\PY@tc##1{\textcolor[rgb]{0.25,0.44,0.63}{##1}}}
\expandafter\def\csname PY@tok@ow\endcsname{\let\PY@bf=\textbf\def\PY@tc##1{\textcolor[rgb]{0.00,0.44,0.13}{##1}}}
\expandafter\def\csname PY@tok@sx\endcsname{\def\PY@tc##1{\textcolor[rgb]{0.78,0.36,0.04}{##1}}}
\expandafter\def\csname PY@tok@bp\endcsname{\def\PY@tc##1{\textcolor[rgb]{0.00,0.44,0.13}{##1}}}
\expandafter\def\csname PY@tok@c1\endcsname{\let\PY@it=\textit\def\PY@tc##1{\textcolor[rgb]{0.25,0.50,0.56}{##1}}}
\expandafter\def\csname PY@tok@kc\endcsname{\let\PY@bf=\textbf\def\PY@tc##1{\textcolor[rgb]{0.00,0.44,0.13}{##1}}}
\expandafter\def\csname PY@tok@c\endcsname{\let\PY@it=\textit\def\PY@tc##1{\textcolor[rgb]{0.25,0.50,0.56}{##1}}}
\expandafter\def\csname PY@tok@mf\endcsname{\def\PY@tc##1{\textcolor[rgb]{0.13,0.50,0.31}{##1}}}
\expandafter\def\csname PY@tok@err\endcsname{\def\PY@bc##1{\setlength{\fboxsep}{0pt}\fcolorbox[rgb]{1.00,0.00,0.00}{1,1,1}{\strut ##1}}}
\expandafter\def\csname PY@tok@kd\endcsname{\let\PY@bf=\textbf\def\PY@tc##1{\textcolor[rgb]{0.00,0.44,0.13}{##1}}}
\expandafter\def\csname PY@tok@ss\endcsname{\def\PY@tc##1{\textcolor[rgb]{0.32,0.47,0.09}{##1}}}
\expandafter\def\csname PY@tok@sr\endcsname{\def\PY@tc##1{\textcolor[rgb]{0.14,0.33,0.53}{##1}}}
\expandafter\def\csname PY@tok@mo\endcsname{\def\PY@tc##1{\textcolor[rgb]{0.13,0.50,0.31}{##1}}}
\expandafter\def\csname PY@tok@mi\endcsname{\def\PY@tc##1{\textcolor[rgb]{0.13,0.50,0.31}{##1}}}
\expandafter\def\csname PY@tok@kn\endcsname{\let\PY@bf=\textbf\def\PY@tc##1{\textcolor[rgb]{0.00,0.44,0.13}{##1}}}
\expandafter\def\csname PY@tok@o\endcsname{\def\PY@tc##1{\textcolor[rgb]{0.40,0.40,0.40}{##1}}}
\expandafter\def\csname PY@tok@kr\endcsname{\let\PY@bf=\textbf\def\PY@tc##1{\textcolor[rgb]{0.00,0.44,0.13}{##1}}}
\expandafter\def\csname PY@tok@s\endcsname{\def\PY@tc##1{\textcolor[rgb]{0.25,0.44,0.63}{##1}}}
\expandafter\def\csname PY@tok@kp\endcsname{\def\PY@tc##1{\textcolor[rgb]{0.00,0.44,0.13}{##1}}}
\expandafter\def\csname PY@tok@w\endcsname{\def\PY@tc##1{\textcolor[rgb]{0.73,0.73,0.73}{##1}}}
\expandafter\def\csname PY@tok@kt\endcsname{\def\PY@tc##1{\textcolor[rgb]{0.56,0.13,0.00}{##1}}}
\expandafter\def\csname PY@tok@sc\endcsname{\def\PY@tc##1{\textcolor[rgb]{0.25,0.44,0.63}{##1}}}
\expandafter\def\csname PY@tok@sb\endcsname{\def\PY@tc##1{\textcolor[rgb]{0.25,0.44,0.63}{##1}}}
\expandafter\def\csname PY@tok@k\endcsname{\let\PY@bf=\textbf\def\PY@tc##1{\textcolor[rgb]{0.00,0.44,0.13}{##1}}}
\expandafter\def\csname PY@tok@se\endcsname{\let\PY@bf=\textbf\def\PY@tc##1{\textcolor[rgb]{0.25,0.44,0.63}{##1}}}
\expandafter\def\csname PY@tok@sd\endcsname{\let\PY@it=\textit\def\PY@tc##1{\textcolor[rgb]{0.25,0.44,0.63}{##1}}}

\makeatother

\providecommand*{\DUrole}[2]{%
  \ifcsname DUrole#1\endcsname%
    \csname DUrole#1\endcsname{#2}%
  \else
    \ifcsname docutilsrole#1\endcsname%
      \csname docutilsrole#1\endcsname{#2}%
    \else%
      #2%
    \fi%
  \fi%
}

\ifthenelse{\isundefined{\hypersetup}}{
  \usepackage[colorlinks=true,linkcolor=blue,urlcolor=blue]{hyperref}
  \urlstyle{same} 
}{}

\begin{document}
\newcounter{footnotecounter}\title{CATOS: Computer Aided Training/Observing System}\author{Jinook Oh$^{\setcounter{footnotecounter}{1}\fnsymbol{footnotecounter}\setcounter{footnotecounter}{2}\fnsymbol{footnotecounter}}$%
          \setcounter{footnotecounter}{1}\thanks{\fnsymbol{footnotecounter} %
          Corresponding author: \protect\href{mailto:jinook.oh@univie.ac.at}{jinook.oh@univie.ac.at}}\setcounter{footnotecounter}{2}\thanks{\fnsymbol{footnotecounter} Cognitive Biology Dept., University of Vienna}\thanks{%

          \noindent%
          Copyright\,\copyright\,2014 Jinook Oh. This is an open-access article distributed under the terms of the Creative Commons Attribution License, which permits unrestricted use, distribution, and reproduction in any medium, provided the original author and source are credited. http://creativecommons.org/licenses/by/3.0/%
        }}\maketitle
          \renewcommand{\leftmark}{PROC. OF THE 6th EUR. CONF. ON PYTHON IN SCIENCE (EUROSCIPY 2013)}
          \renewcommand{\rightmark}{CATOS: COMPUTER AIDED TRAINING/OBSERVING SYSTEM}

\setcounter{page}{15}
\newcommand*{\docutilsroleref}{\ref}
\newcommand*{\docutilsrolelabel}{\label}
\AtEndDocument{\cleardoublepage}
\begin{abstract}In animal behavioral biology, there are several cases in which an autonomous observing/training system would be useful. 1) Observation of certain species continuously, or for documenting specific events, which happen irregularly; 2) Longterm intensive training of animals in preparation for behavioral experiments; and 3) Training and testing of animals without human interference, to eliminate potential cues and biases induced by humans. The primary goal of this study is to build a system named CATOS (Computer Aided Training/Observing System) that could be used in the above situations. As a proof of concept, the system was built and tested in a pilot experiment, in which cats were trained to press three buttons differently in response to three different sounds (human speech) to receive food rewards. The system was built in use for about 6 months, successfully training two cats. One cat learned to press a particular button, out of three buttons, to obtain the food reward with over 70 percent correctness.\end{abstract}

\section{Introduction%
  \label{introduction}%
}

It is often the case in animal behavioral biology that a large amount of human resources, time, and data storage (such as video recordings) are required in animal observation and training. Some representative examples of these cases are:%
\begin{itemize}

\item 

Observation of certain species continuously or monitoring for specific events, which occur irregularly, when behavior of certain species during any time period or specific time period, such as nocturnal behaviors, are investigated.
\item 

Certain experiments require a prolonged training period, sometimes over a year. This type of experiment requires reliable responses, which may not correspond to usual behavior patterns, from animals in tasks. Therefore, training may require a long period of time until the subject is ready to be tested. Additionally, long periods of human supervised training can introduce unintended cues and biases for animals.
\end{itemize}

In the first case, an autonomous system for observing animals can save human resources and reduce the amount of data storage. The reduced amount of data can also conserve other types of human resources such as investigation and maintenance of large-scale data. There have been attempts to build autonomous observing or surveillance systems in the fields of biology, such as Kritzler et al. \cite{Kri08}’s work, and security systems, such as Belloto et al. \cite{Bel09}, Vallejo et al. \cite{Val09}, for instance. There are also commercial products for surveillance systems with various degrees of automation, or incorporating artificial intelligence. However, the intelligence of each system is case-specific and it is difficult to apply these specific systems to novel situations without considerable adjustments. In the second case, an autonomous system for prolonged, intensive training can also save human resources and eliminate potential cues and biases caused by humans. Training with an autonomous system is an extension of traditional operant conditioning chambers and many modern and elaborated versions have been developed and used, such as in Markham et al. \cite{Mar96}, Takemoto et al. \cite{Tak11}, Kangas et al. \cite{Kan12}, Steurer et al. \cite{Ste12}, and Fagot \& Bonte \cite{Fag09}. However, many of the previous devices use commercial software. Also, they do not possess the observational features developed in the current project. It would be useful to have an open-source, relatively low-budget, and modularized system which could be customized for the observation, training and the experimentation on animal subjects of various species. CATOS, the system built in the present study, fulfills these necessities. The difference between the previous systems and CATOS (Computer Aided Training/Observing System) in the present work is that the animals do not have to be captured or transported to a separated space at a specific time in order to be trained. The disadvantages of separating animals (e.g., primates) are well-known, and include stress on animals separated from their group or moved from their usual confines, the risky catching procedure for both animal and human (cf. Fagot \& Bonte \cite{Fag09}). Similar arguments apply to most animal species, especially when they are social. The automatic learning device for monkeys (ALDM) described in Fagot \& Bonte \cite{Fag09} is very similar to the trainer aspect of CATOS described in the present work, but CATOS is different in following features. First of all, it aimed to be open-source based and more modular so that it can be more easily adjusted and adopted to different species and experiments. Another feature is that CATOS is equipped with various observational features, including visual and auditory recording and recognition through video camera and microphone, which make the system able to interact with the subjects, such as reacting immediately to a subject with a motion detection from a camera or a sound recognition from a microphone.
CATOS should offer the following advantages.%
\begin{itemize}

\item 

The system should be flexible in terms of its adjustability and the extendibility to various projects and species. The software should be open-source, and both software and hardware components should be modularized as much as possible, thus the system reassembly for researchers in animal behavioral biology is practical.
\item 

The system should have various observational features applicable to a broad range of animal species and observational purposes.
\item 

The system should perform continuous monitoring, and it should record video and/or sound only when a set of particular conditions is fulfilled. This would reduce the amount of data produced during the procedure.
\item 

The system should have actuators to react in certain situations, which allows it to act as a trainer/experimenter. The human trainer/experimenter designs the procedure by adjusting parameters and modules, but the actual performance should be done by the system. In this way, the system could help reducing the amount of time required for training, and eliminating cues/biases which might be induced by the human interferences.
\item 

With this system, the animal should not have to be transported to a certain space, or separated from its group, for training. The animals should be able to choose when to start a trial on their own.
\end{itemize}

Two CATOS prototypes have been built during this study. The first build of CATOS has 3 pushbuttons as a main input device for cats and the second build has a touch-screen as a main input device. The first build was an initial attempt to build and test such a system. The second build is the final product of the study. The basic structures of these two builds are more or less the same. The differences are that the second version has improved functions and it uses the touch-screen instead of pushbuttons. The first build of CATOS was tested with domestic cats (Felis catus) to train them to press three different buttons differently depending on the auditory stimuli (three different human speech sounds). The final goal of this training is to investigate human speech perception in cats. There is no doubt in that many animal species can recognize some words in human speech. The examples of speech perception in dogs and chimpanzees can be found in the work of Kaminski et al. \cite{Kam04} and Heimbauer et al. \cite{Hei11} respectively. In some cases, animals can even properly produce words with specific purposes. An example of speech perception and production in a parrot can be found in the work of Pepperberg \cite{Pep87}. Despite these findings, there is ongoing debate about whether the same perceptual mechanisms are used in speech recognition by humans and animals (Fitch \cite{Fit11}). To investigate this issue, animals have to be trained to show different and reliable responses to different human speech sounds. Then, we can test which features of human speech are necessary for different animal species to understand it. Thus, the final aim of the training in this study would be to obtain cats showing different responses to different human speech sounds with statistical significance (over 75 percent). Before reaching this final goal, several smaller steps and goals are required.

\section{Brief description of CATOS (Computer Aided Training/Observing System)%
  \label{brief-description-of-catos-computer-aided-training-observing-system}%
}

The overall system is composed of a combination of software and hardware components. The software components are mainly composed of the Python script named as ’AA.<version>.py’ and the program for the microcontroller. The ’AA’ runs all of the necessary processes and communicates with the microcontroller program. The microcontroller program operates sensors and actuators as it communicates with the ’AA’ program. The hardware components are composed of various devices, some of which are directly connected to the computer via USB cables. Some other devices only have GPIO (General Purpose Input Output) pins; therefore they are connected to the microcontroller. The microcontroller itself is connected to the computer via a USB cable. The hardware devices, which are directly connected via USB cables, can be accessed using various software modules, which are imported into the ’AA’ program. The access to other devices only using GPIO pins is performed in the microcontroller and the ’AA’ program simply communicates with the microcontroller program via a serial connection for sending commands to actuators and receiving values from sensors.\begin{figure}[]\noindent\makebox[\columnwidth][c]{\includegraphics[width=\columnwidth]{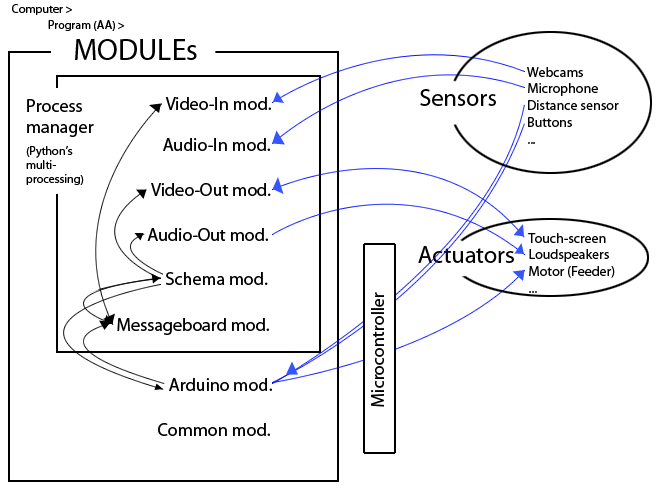}}
\caption{Overall system diagram. \DUrole{label}{dgSystem}}
\end{figure}

The software for this system is called AA (Agent for Animals). This software was build with helps of many external libraries such as OpenCV \cite{Bra00}, and NumPy/SciPy \cite{Jon01}. Once it starts, seven processes were launched using multiprocessing package of Python and it runs until the user terminates the program. The multiprocessing was used because the heavy calculation for image processing from multiple webcams were concerned. The number of processes can be changed as some of them can be turned on or off. These processes include a video-in process for each camera, a video-out process, an audio-in process, an audio-out process, a schema process, and a message-board process. Figure \DUrole{ref}{dgSystem}. Even though some of these processes have quite simple tasks, they were separated in order to prevent them from interfering with each other and/or becoming the bottleneck. The system has to process the visual, auditory, and other sensory and motor information simultaneously to recognize the change of the environment and respond to it properly. The output data such as captured video input images, recorded WAV files, movement-records, CSV files for trial results, and the log file are temporarily stored in the ’output’ folder. After the daily session is finished, all of these output files go through an archiving process which can include, but is not restricted to, generating movies, generating images with the movement analysis, labeling sound files, and moving different types of files into the categorized subfolders of an archiving folder named with a timestamp.\begin{figure}[]\noindent\makebox[\columnwidth][c]{\includegraphics[width=\columnwidth]{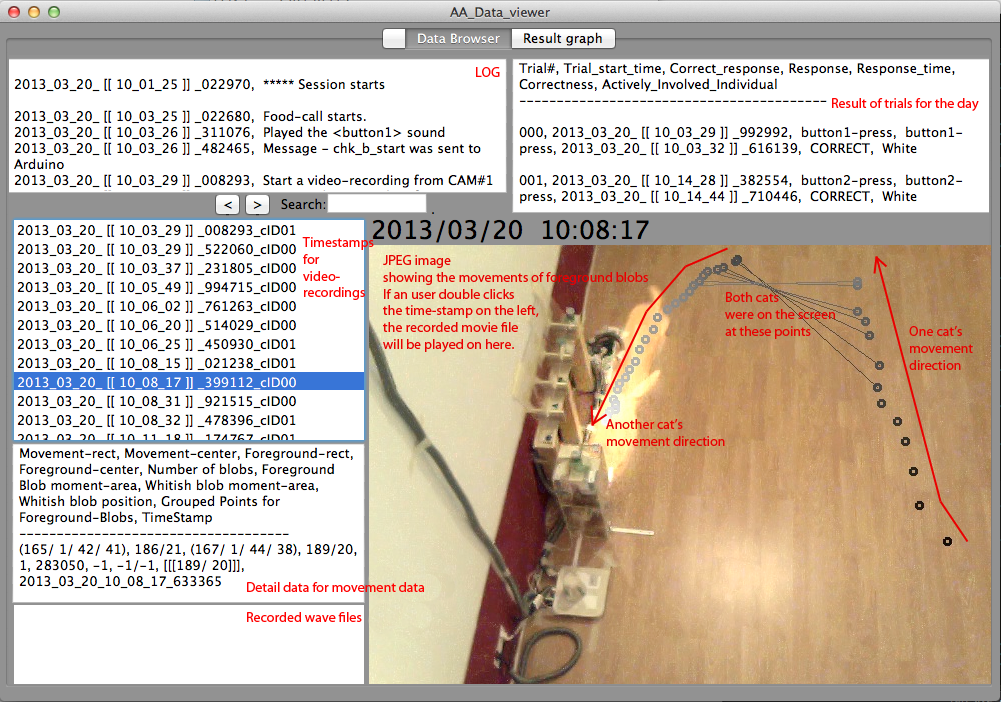}}
\caption{AA\_DataViewer \DUrole{label}{AADataViewer}}
\end{figure}

Besides combining all the above modules and implementing some common functions, one more Python program was implemented to facilitate the process of analyzing the recorded data. The program is called ”AA DataViewer”, which is based on wxPython GUI toolkit and Matplotlib \cite{Hun07} for drawing graphs. Figure \DUrole{ref}{AADataViewer}. It loads the log file, the result CSV (comma separated values) file containing the results of the trial, the movement-record CSV files, the MP4 movie files, and the WAV files from one folder containing all data collected for one session (day). For each video clip, there is a JPEG image showing the movements of the blobs. The circles in the image represent the positions of the blobs and their color represents the time-flow, with the black corresponding to the beginning of the movie, and the white to the end of the movie. A line connecting multiple circles means that those blobs occurred at the same time. Another feature of this program is its ability to generate a graph with selected sessions. In the ’archive’ folder, there are sub-folders, each of which contains all the data for a session. When the ’select sessions’ button is clicked, a pop-up window appears for selecting multiple folders. The result data from these selected sub-folders of ’archive’ folder is drawn as a graph using Matplotlib \cite{Hun07}. By visualizing the data for certain period, it helps the trainer or experimenter quickly assess the current status of the training procedure.\begin{figure}[]\noindent\makebox[\columnwidth][c]{\includegraphics[width=\columnwidth]{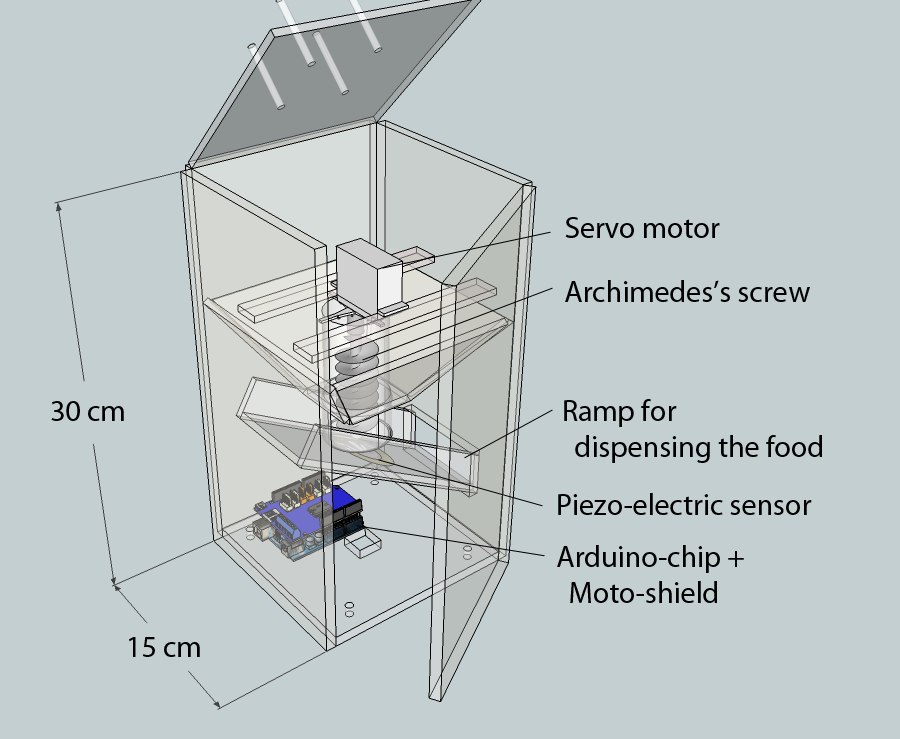}}
\caption{Automatic feeder \DUrole{label}{feeder}}
\end{figure}

The two feeders used in this study is a device mainly comprising the Arduino microcontroller; (refer \url{http://www.arduino.cc/}), a motor-shield for the microcontroller, a servomotor, and a frame encasing the whole feeder. Both Feeder variants work in a similar way, by rotating the servomotor by a certain number of degrees, although the second feeder shows better performance in terms of consistent amount of food released, due to the usage of an Archimedes’ screw. Initially, an estimate of the amount of food left in the food container was obtained using an IR distance sensor, but this feature was discarded in the second build since the distance information from the IR sensor was not accurate enough for this application. The second feeder confirms the emission of a food reward via the piezoelectric sensor, which is positioned right below the Archimedes’ screw. Figure \DUrole{ref}{feeder}.\begin{figure}[]\noindent\makebox[\columnwidth][c]{\includegraphics[width=\columnwidth]{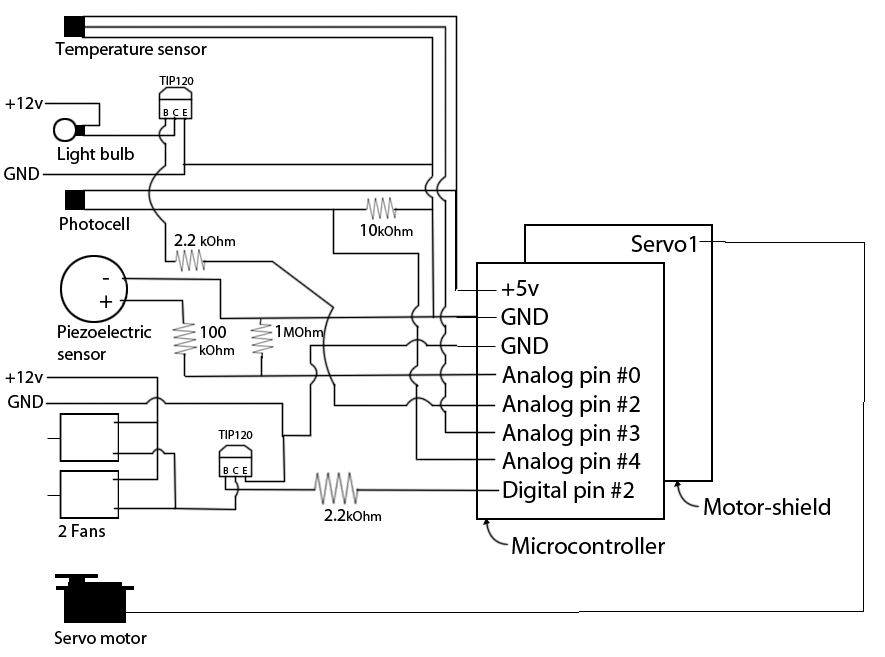}}
\caption{Circuit with a microcontroller \DUrole{label}{circuit}}
\end{figure}

Communication between the Arduino chip and the main computer was accomplished by using the Arduino module of the ’AA’ program.
In the circuit, Figure \DUrole{ref}{circuit},%
\begin{itemize}

\item 

The temperature sensor measures the temperature inside of the protective wooden platform.
\item 

The photocell sensor measures the ambient light level.
\item 

The light bulb can be turned on when the photocell sensor indicates the ambient light level is below a user-defined threshold.
\item 

Two fans are turned on when the temperature sensor indicates the temperature is too high in the platform.
\item 

The piezoelectric sensor is read while the servomotor is actuating, in order to confirm the occurrence of the food reward. This sensor reading is required because occasionally the food dispensing fails due to the combination of the short motor activation time (<0.5 seconds) and the shape of the dry food pieces (which can fit into other pieces easily and then fail to emerge).
\item 

The servomotor is responsible for the food dispense by turning the Archimedes’ screw back and forth.
\end{itemize}

\section{Results of building CATOS and its testing on 2 domesticated cats%
  \label{results-of-building-catos-and-its-testing-on-2-domesticated-cats}%
}

The hardware and software were built and tested. The software is available at \url{https://github.com/jinook0707/CATOS_alpha} with GNU General Public License, version 3. Both hardware and software are curretnly in its alpha stage. Although its potential to be used to train and test animal cognition was tested and its usage seemed promising to save human resources in certain situations, both hardware and software should be developed further to be practically used for experimenting animal cognition.

The two web-cams observed the experimental area for 8 to 12 hours per day for about 5 months (from the middle of October 2012 to the middle of March 2013). The movement records, MP4 movie files, JPEG image files, and WAV sound files generated during this period took 37.35 Giga bytes of storage. To obtain a rough idea of the degree of reduction in data storage that was achieved using the system, the number of recorded frames in the video recording was assessed. Data for 15 days were taken to calculate it. The total observation period was 406138 seconds, corresponding to 112.8 hours. The number of frames recorded was 206024 and the average FPS(Frame Per Second) was 7.5, therefore, approximately, the video recordings were stored for 27470 seconds (=7.6 hours), which is about 6.7 percent of entire observation period. These specific numbers are not very meaningful since they can fluctuate with the increase or decrease of the subject’s movements, but the point is that the most of the meaningless recordings were successfully filtered out by CATOS.

Human presence during session is not necessary. Data transfer from one computer to another, maintenance, or modification of the system requires human interaction, but no time and effort is required concerning the training and testing sessions. Because no one attends the sessions, a periodic analysis of the animal’s performance with the system is required. A simple assessment of how much food the animals took, or more specifically, how many correct and incorrect trials occurred, can be done quickly since this information is already stored in result CSV file displaying the number of correct and incorrect trials generated with timestamps at the end of each session. Also, the data-viewer utility program displays all the timestamps and its JPEG image, which presents a brief report on the movement detected in the recorded video-clip. Thus, simply browsing the JPEG images is often enough to assess the session. If it is not enough, then one can obtain a more detailed assessment by playing the video-clips recorded around the trial times.\begin{figure}[]\noindent\makebox[\columnwidth][c]{\includegraphics[width=\columnwidth]{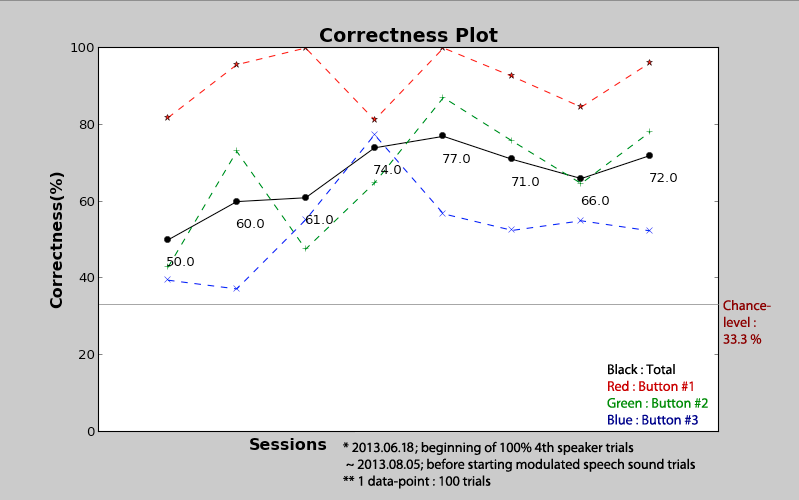}}
\caption{Recent performance of the trained cat on three human speech discrimination task. \DUrole{label}{recPerf}}
\end{figure}

Two domesticated cats were trained for testing the system. Both cats learned that approaching the feeder on a playback sound could lead to a food reward. Then one cat further learned that pressing one out of three buttons could lead to a food reward. The training of the association between three different sound stimuli and three different buttons is an ongoing process. The most recent performance data Figure \DUrole{ref}{recPerf} shows over 70 percent of overall performance and also the performance on each button is significantly higher than 33.3 percent of chance level.

\end{document}